# Ordering in Red Abalone Nacre

Rebecca A. Metzler[1], Dong Zhou[1], Mike Abrecht[2], Susan N. Coppersmith[1],

and P.U.P.A. Gilbert[1,*]

[1] *Department of Physics, University of Wisconsin, Madison, WI 53706, USA.*

[2] *Synchrotron Radiation Center, 3731 Schneider Drive, Stoughton, WI 53589, USA.*

* *Previously publishing as Gelsomina De Stasio. Corresponding author: pupa@physics.wisc.edu*

**Red abalone nacre is an intensely studied biomineral, and yet its formation mechanism remains poorly understood.[1-5] Here we report quantitative measurements probing the degree of order of the aragonite tablets in nacre, and show that order develops over a distance of about 50 microns. These observations indicate that the orientational order of aragonite tablets in nacre is established gradually and dynamically, and we show that a model of controlled assembly based on suppression of the crystal growth rate along a specific direction, when growth is confined in a layered structure, yields a tablet pattern consistent with those revealed by detailed experimental measurements[3]. This work provides strong evidence that the organism's control of crystal orientation in nacre occurs via regulation of crystal nucleation and growth as opposed to direct templation[6] or heteroepitaxial growth[4] on organic molecules on the organic matrix sheets.[5,7,8]**

Nacre, or mother-of-pearl, is widely studied because of its self-assembled, efficient and accurately ordered architecture, its toughness, and its fascinating and poorly understood formation mechanisms.[1] Nacre is a composite of layered 400-nm thick aragonite crystalline tablets oriented with their (001) crystal axes within ±11° of the normal to the layer plane[7] and stacked irregularly[3], and 30-nm thick organic matrix sheets.[9] Aragonite, a hard but brittle orthorhombic $CaCO_3$ polymorph, accounts for 95% of nacre's mass, yet nacre is 3000 times tougher than aragonite.[10] No synthetic composites outperform their components by such large factors. It is therefore of great interest to understand nacre formation.

We present here x-ray photoelectron emission spectromicroscopy (X-PEEM) data on red abalone shells that yield unprecedented detailed and quantitative information about the location, size and orientation of individual tablets with sub-micron spatial resolution, a large field of view including hundreds of tablets, straightforward sample preparation, and negligible radiation damage. The data exhibit a systematic spatial variation, indicating that the aragonite tablets in nacre order dynamically, and are consistent with a theoretical model in which aragonite crystal layers are nucleated sequentially in the presence of confining matrix sheets and grow epitaxially on aragonite crystals in layers below, with the growth rate of aragonite crystals in the (001) crystal direction lower than in the ab-plane.

Illuminating polycrystalline samples with polarized soft-x-rays generates a spectroscopic effect known as x-ray linear dichroism[11,12] in x-ray absorption near-edge structure (XANES) spectra.[13] This effect is widely studied in man-made liquid crystals,[12] chemisorbed[11] and magnetic materials, [14,15] occasionally in geologic minerals,[16,17] and was recently discovered in nacre.[3] When XANES spectroscopy is combined with X-PEEM imaging, the linear dichroism generates imaging contrast between differently oriented adjacent microcrystals, denoted polarization-dependent imaging contrast (PIC).[3,18] Because of aragonite's hexagonal symmetry in the ab-plane, the contrast arises from differences in the crystals' c-axes orientation *(3,13)*.[3,13] In this work we analyzed eleven samples from four different red abalone shells, approximately 10-cm long, with the X-PEEM Spectromicroscope for PHotoelectron Imaging of Nanostructures with X-rays (SPHINX).[18] The shells were cut, embedded, polished and Pt coated as described previously.[5]

We characterize quantitatively tablet size, tablet crystal orientation, and stacking direction; they all show greater disorder near the boundary between the nacre and prismatic layers, with order evolving over a length scale of approximately 50 microns. Figure 1 shows a series of O maps acquired with SPHINX at and away from the prismatic boundary. In Figure 1 adjacent tablets exhibit strikingly strong contrast, as shown by the different gray levels, due to different orientations of adjacent crystalline tablets in each nacre layer.[17]

As can be seen in Figure 1, the contrast is greatest near the nacre-prismatic boundary, and diminishes as the distance from the boundary increases. Figure 2 reports the magnitude of the variation of PIC at each distance from the boundary. The contrast arising from c-axis misorientation of tablet crystals is largest near the nacre-prismatic boundary, and decreases moving away from the boundary, reaching its steady-state value at distances of 50 μm. Low but still distinct PIC is observed across the whole 2-mm thickness of nacre.

Supplementary Table 1 reports PIC range data for one shell up to 50 μm from the boundary. The disorder in tablet crystal orientation (Figure 2) as revealed by the variability in the gray scale of the image and in tablet widths (Supplementary Figure 1) evolve systematically as one moves away from the prismatic boundary. The systematic evolution of the degree of order is incompatible with a mechanism in which tablet crystal orientation is determined entirely by direct templation by the organic matrix sheets or acidic macromolecules on the sheets, because such mechanism would not yield evolution of the degree of order as one moves away from the prismatic boundary. Our results are consistent with previous synchrotron x-ray micro-diffraction observations reporting that nacre tablets in red abalone are less co-oriented near the nacre-prismatic boundary, and become more co-oriented and ordered within 100 μm of the boundary.[19] PIC resolves orientation differences between individual neighboring tablets in a large field of view, providing statistical information that is a basis for comparison with theoretical modeling.

Figure 3A, a PIC image taken near the nacre-prismatic boundary, has stacks of co-oriented tablets with the same gray level, consistent with the idea that mineral bridges[20] through porous organic nucleation sites[21,22,3] conserve the orientation of crystals in successive layers. The stacks of co-oriented tablets have finite height, so it is plausible that only a small fraction of nucleation sites nucleate new crystal orientations.

The main and new assumption we make in our model is that:

• The growth rate along the (001) direction (the c-axis direction) of aragonite tablets in nacre is slower than the growth rates along the a- and b-axes. This assumption is consistent with reports that nacre proteins suppress the growth rate of carbonates along (001).[23,24,25]

The other, more conventional, assumptions are:

- Organic matrix sheets confine the growth of the aragonite tablets.[2,8]
- There is a single nucleation site per nacre tablet.[2,8]
- Nucleation sites are independently and randomly distributed on organic matrix sheets.[3]
- Tablets in a given layer grow until they reach confluence.[26]
- A tablet in a given layer is highly probable to have the same orientation as the tablet directly below its nucleation site. [20]
- The c-axis orientations of the aragonite tablets in the first layer are not aligned. This assumption is consistent with previous observations,[19] and further supported by the experimental data presented here.
- In the calculations shown here, growth of a given layer is completed before the next layer is nucleated. This limit is adopted purely for simplicity; more realistic columnar growth[27] actually increases the probability that tablets with high growth velocity preferentially reach the nucleation sites in the overlying matrix sheet, which enhances the orientational ordering via the mechanism we describe.

These assumptions yield the simulated model configuration shown in Figure 3B, which is strikingly similar to the experimental data in Figure 3A. The spread in the growth speeds of different crystal orientations determines the length over which the ordering evolves, and the fraction of tablets that fail to adopt the crystal orientation of the tablets below their nucleation sites determines the ultimate degree of order far from the prismatic boundary (in Fig. 3B, the spread in growth velocities is ~20%, and 2.8% of tablets are misaligned with respect to the underlying ones). Supplementary Figure 2 shows that both the model and red abalone nacre have occasional configurations with tablet widths that are significantly larger than in typical configurations. The growth model's behavior and its relations to models studied in the context of population biology are discussed further in the Supplementary Information.

The remarkable resemblance of the experimentally observed tablet orientation patterns and the configurations yielded by the model provide strong evidence that the growth rate of aragonite crystals in nacre is greatest when the c-axes are oriented perpendicular to the organic matrix sheets. We now show why suppression of the rate of aragonite crystal growth along (001) gives rise to tablet growth rates with this property.

Because the nacre tablets are constrained by organic matrix sheets, decreasing the growth rate along (001) until it is slower than along the a- and b-axes causes the tablets with (001) axis perpendicular to the matrix sheets to grow the fastest. (This mechanism assumes that an "ordering protein" is available to suppress the aragonite growth rate along (001) in the inter-sheet space or gel in which aragonite tablets grow.[28,29] In this scenario the "ordering protein" acts stereochemically.[23] The organic matrix sheets play a critical role – in their absence, crystals align with their fastest-growing axes parallel to the growth direction, as in eggshells.[30] Figure 4 illustrates the mechanism; when the crystal is oriented with the c-axis perpendicular to the sheets, then it can grow in its fast a and b directions without running into the organic matrix sheets.

We investigated the spatial distribution of proteins in nacre at and near the boundary, by mapping nitrogen. The results of Supplementary Figure 5 show greater concentration of proteins in nacre than in the prismatic layer. The total protein concentration in nacre does not exhibit a gradient but is constant at and away from the boundary.

In summary, we have demonstrated that the complex pattern of aragonite tablet orientations in nacre provides evidence that the crystal orientation order of the tablets is induced by a dynamical mechanism implemented by an "ordering protein" that suppresses the rate of aragonite crystal growth along (001). The ordering mechanism presented here contrasts with the common hypothesis that aragonite order is the result of templation by or heteroepitaxial growth on organic molecules on the matrix sheets. Direct and abrupt templation from the organic matrix sheets does not account for the characteristic pattern of tablet orientations, nor for the gradual increase of order observed moving away from the prismatic boundary. The dynamic mechanism of growth control by an "ordering protein" described here provides a realistic theoretical growth model, in good agreement with experimental data.

**Acknowledgements**

This work was supported by NSF awards PHY-0523905 and CHE-0613972, and UW-Graduate School Vilas Award to PUPAG, and NSF award DMR-0209630 to SNC. The experiments were performed at the UW-SRC, supported by NSF award DMR-0537588.

**Supporting Online Material**

Detailed quantitative characterization of the evolution of order observed in our experiments (Supplementary Table 1, Supplementary Figure 1)

Demonstration that sample-to-sample variability in the model is consistent with the variability observed in the experimental data (Supplementary Figure 2)

Discussion of the theoretical model (Supplementary Text Section 1)

Numerical results from the model (Supplementary Text Section 1a, Supplementary Figures 3 and 4)

Results from analytic theory (Supplementary Text Section 1b)

Oxygen, nitrogen, and calcium maps (Supplementary Figure 5)

Tablet stacking direction versus distance from the prismatic-nacre boundary (Supplementary Figure 6).

Supplementary References

Figure 1

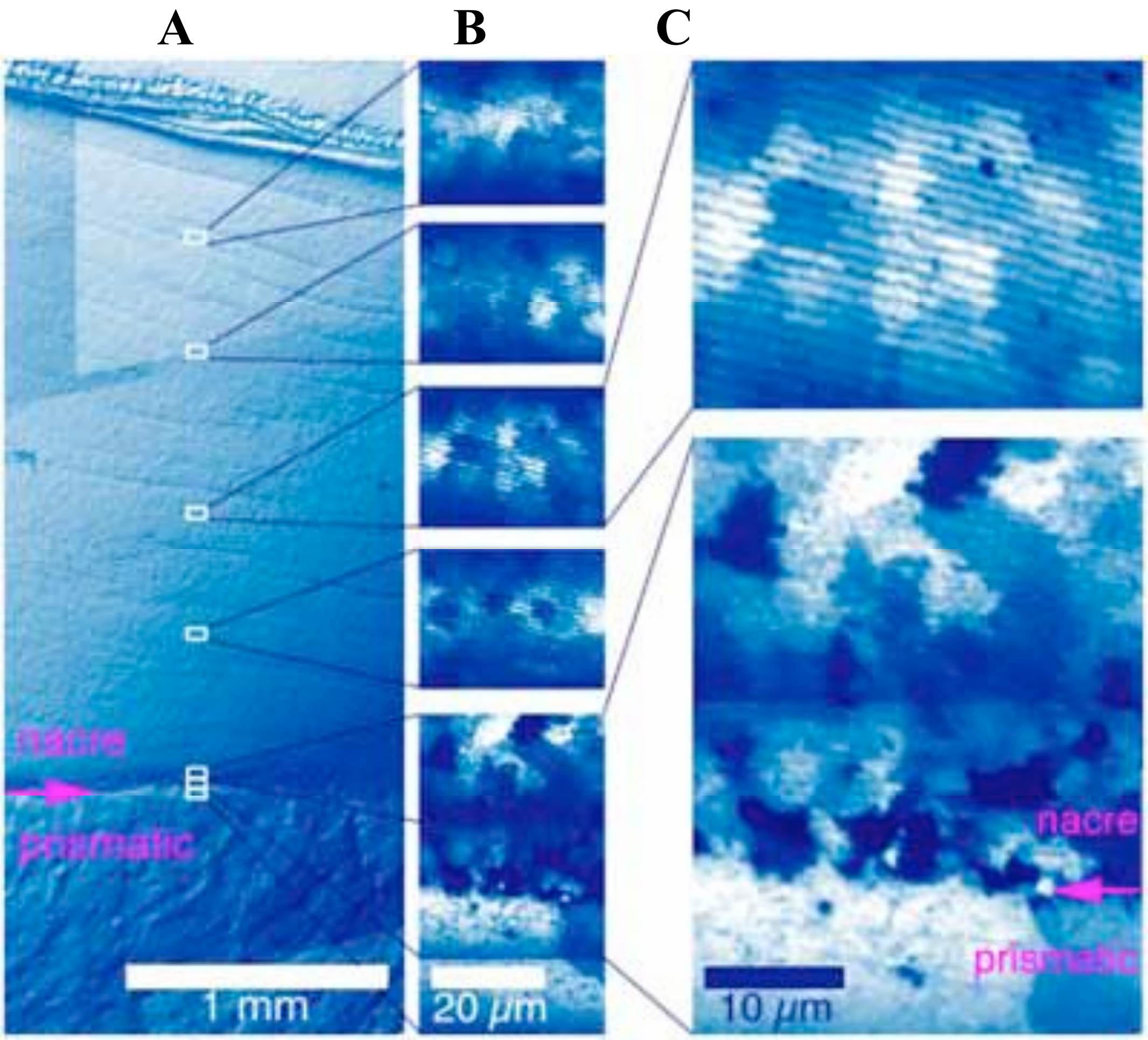

Figure 1. (**A**) Visible light microscope composite image of a polished section of red abalone shell, acquired in differential interference contrast (Nomarski) mode. The rectangles outline the areas in which the O maps in (B) were acquired with SPHINX, while the magenta arrow indicates the nacre prismatic boundary, which runs horizontally across the entire image. (**B**) Series of five X-PEEM polarization-dependent imaging contrast (PIC) oxygen maps, obtained from ratios of image at 534 eV and 518 eV, that is, π*/pre-edge. The pixel intensity in the SPHINX oxygen or carbon PIC maps the angle between the tablet c-axis and the polarization vector. The different gray levels in PIC maps correspond to variations in the intensity of the 534 eV peak, which is the O1s→π* transition of the carbonate CO bonds. (The 290.3 eV peak of C1s→π*, not shown, maps similarly.) The growth direction was 61° from the polarization vector and 60° from the sample normal. These PIC maps demonstrate how the π* PIC contrast diminishes with the distance from the boundary, but persists across the entire nacre thickness. (**C**) Magnified PIC oxygen maps at and away from the boundary (arrow), showing co-oriented columns of tablets, and the contrast among them.

Figure 2

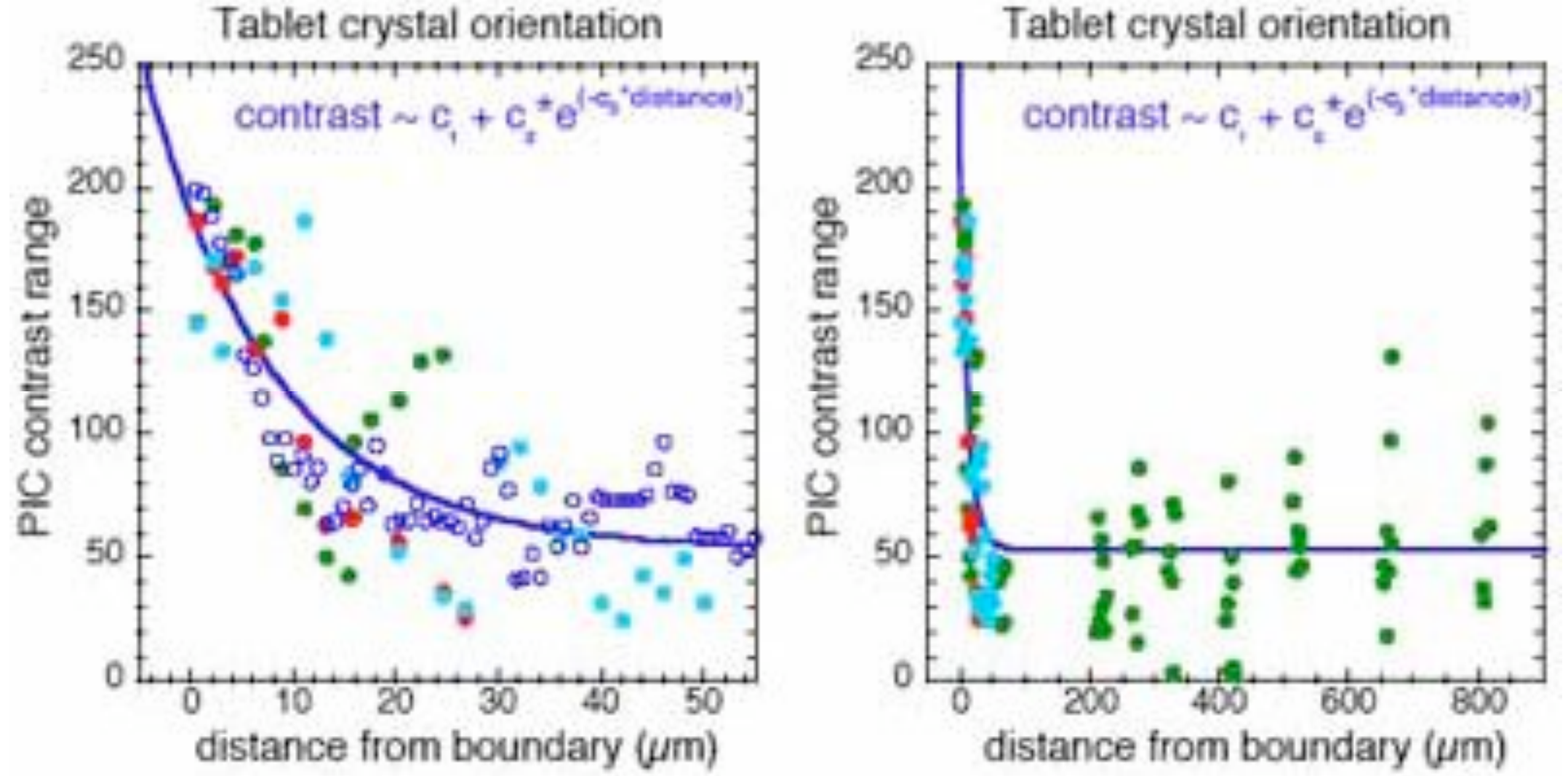

Figure 2. The range of polarization-dependent imaging contrast (PIC) versus distance from the prismatic boundary, in the direction of nacre growth. At each distance, in oxygen PIC maps, we measured the mean gray level in the brightest tablet, in the darkest tablet, and subtracted the latter from the former to obtain the PIC range. (**A**) PIC range from three different red abalone shells, represented by solid colored dots. The contrast range decreases with distance from the boundary, stabilizing at 50 µm. Simulation data from the theoretical model are also reported as open blue circles. (**B**) In one red abalone shell the PIC range was measured up to a distance of 800 µm and found to fluctuate around 50 at distances greater than 50 µm from the boundary. The data exhibit considerable scatter, but are consistent with numerical simulations of our theoretical model (the details of this comparison are in the Supplementary Information). The blue line is a fit of the model data, with fit parameters: $c_1 = 54$; $c_2 = 135$; and $c_3 = 0.08$.

Figure 3

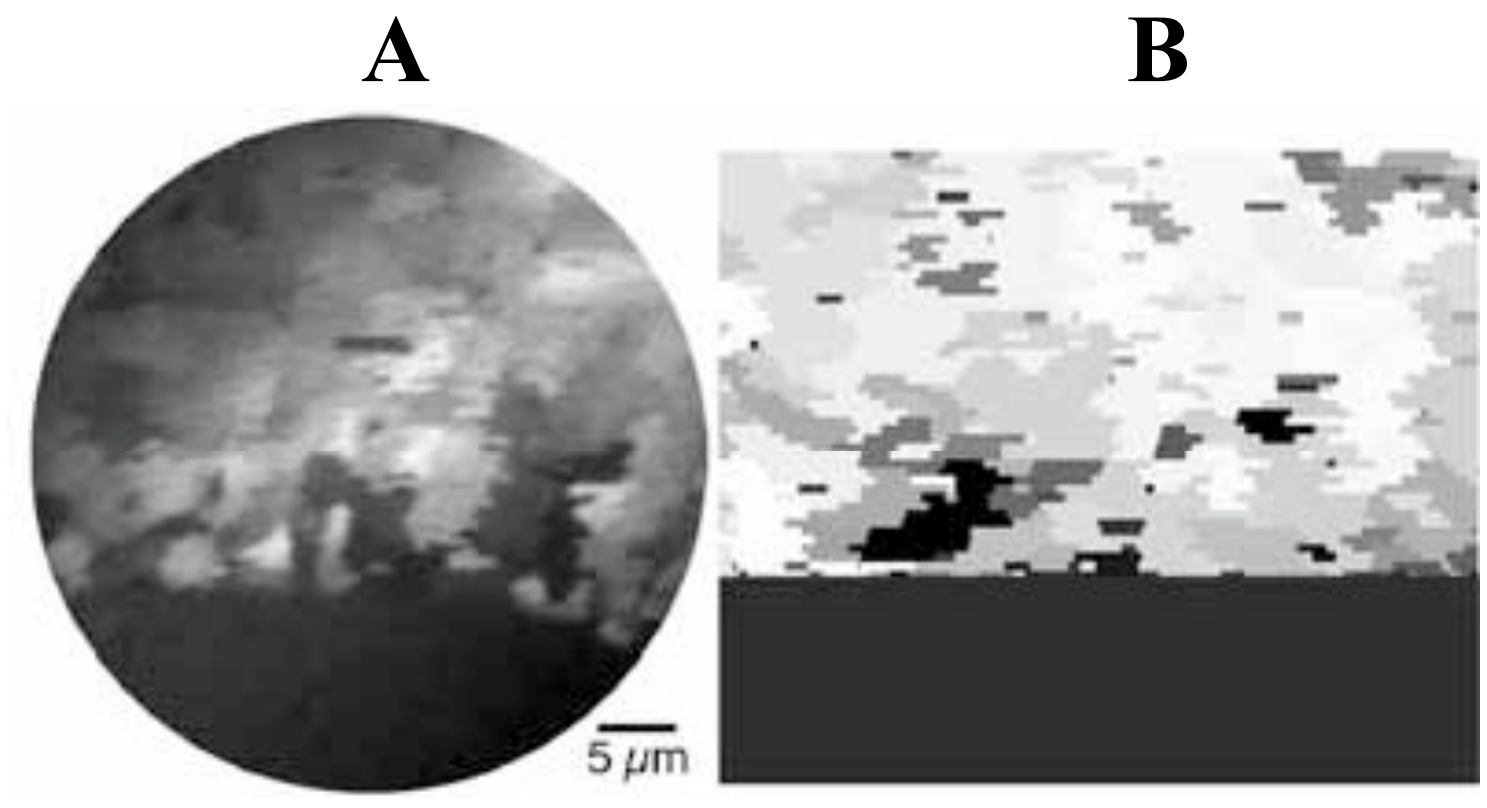

Figure 3. (**A**) Oxygen PIC data showing the pattern of contrast for a region of nacre (top) near the boundary with the prismatic layer (bottom). Stacks of tablets with the same gray level have the same crystal orientation, and the decrease in overall contrast moving away from the nacre-prismatic boundary indicates that tablet c-axis orientations become closer to perpendicular to the organic matrix layers.
(**B**) Simulation results for a model in which each layer grows to completion before the next layer is nucleated, all nucleation sites are randomly distributed and uncorrelated, and each tablet adopts the orientation of the tablet below its nucleation site with probability $1-\varepsilon$. The growth rates of the tablets in the first layer as well as those not co-oriented with the tablet below are chosen uniformly and randomly in the range $[1-\delta/2,\ 1+\delta/2]$. Parameter values are $\varepsilon=0.028$, $\delta=0.2$. A two-dimensional vertical slice through the three-dimensional simulation is shown. Gray levels denote the growth rates of the tablets (light represents higher growth rate). The decrease in disorder moving away from the boundary and the overall morphology are remarkably similar to the experimental data.

Figure 4

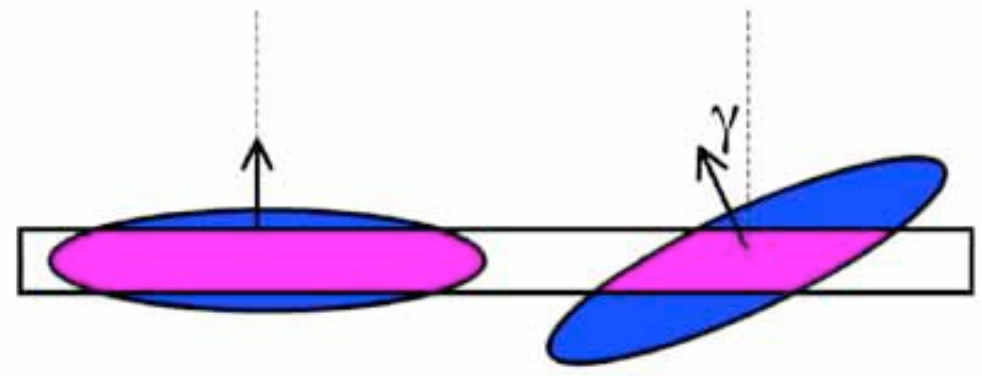

Figure 4. Illustration of why slowing crystal growth along the c-axis in the presence of a confining layered organic matrix yields slower tablet growth. The two ellipses have the same area, but when the minor axis is misoriented from the layer normal, more of the area is outside the layer (shown in blue). When growth is arrested as the crystal runs into the organic matrix sheet, then the amount of tablet that actually grows inside the confined layer (shown in magenta) is greater for a tablet with the c-axis perpendicular to the sheets, that is, parallel to the nacre growth direction. The smaller the angle $\gamma$ the faster the growth.

# Supplementary Information

# Ordering of Crystalline Tablets in Nacre

Rebecca A. Metzler[1], Dong Zhou[1], Mike Abrecht[2], Susan N. Coppersmith[1],

and P.U.P.A. Gilbert[1,*]

[1] *Department of Physics, University of Wisconsin, Madison, WI 53706, USA.*

[2] *Synchrotron Radiation Center, 3731 Schneider Drive, Stoughton, WI 53589, USA.*

* *Previously publishing as Gelsomina De Stasio. Corresponding author:* *pupa@physics.wisc.edu*

This supplement contains:
Detailed quantitative characterization of the evolution of order observed in our experiments (Supplementary Table 1, Supplementary Figure 1)
Demonstration that sample-to-sample variability in the model is consistent with the variability observed in the experimental data (Supplementary Figure 2)
Discussion of the theoretical model (Supplementary Text Section 1)
Numerical results from the model (Supplementary Text Section 1a, Supplementary Figures 3 and 4)
Comparison of model with experimental data (Supplementary Text Section 1b)
Results from analytic theory (Supplementary Text Section 1c)
Oxygen, nitrogen, and calcium maps (Supplementary Figure 5)
Tablet stacking direction versus distance from the prismatic-nacre boundary (Supplementary Figure 6)
Supplementary References

Supplementary Table 1. Quantification of evolution of PIC contrast as a function of distance from the prismatic-nacre boundary.

| **Distance from boundary (μm)** | **Brightest tablet mean gray level** in O map π*/pre-edge | **Darkest tablet mean gray level** in O map π*/pre-edge | **PIC contrast range** (brightest-darkest mean gray levels) |
|---|---|---|---|
| 0.5 | 183 | 38 | 145 |
| 2 | 220 | 49 | 171 |
| 3 | 177 | 43 | 134 |
| 4 | 208 | 43 | 165 |
| 6 | 214 | 46 | 168 |
| 9 | 210 | 55 | 155 |
| 11 | 241 | 54 | 187 |
| 13 | 192 | 53 | 139 |
| 15 | 132 | 48 | 84 |
| 16 | 127 | 45 | 82 |
| 20 | 101 | 48 | 53 |
| 25 | 82 | 47 | 35 |
| 27 | 79 | 49 | 30 |
| 30 | 193 | 104 | 89 |
| 32 | 201 | 106 | 95 |
| 34 | 186 | 107 | 79 |
| 36 | 174 | 114 | 60 |
| 38 | 162 | 103 | 59 |
| 40 | 134 | 102 | 32 |
| 42 | 133 | 108 | 25 |
| 44 | 142 | 99 | 43 |
| 46 | 134 | 98 | 36 |
| 48 | 148 | 98 | 50 |
| 50 | 180 | 108 | 32 |

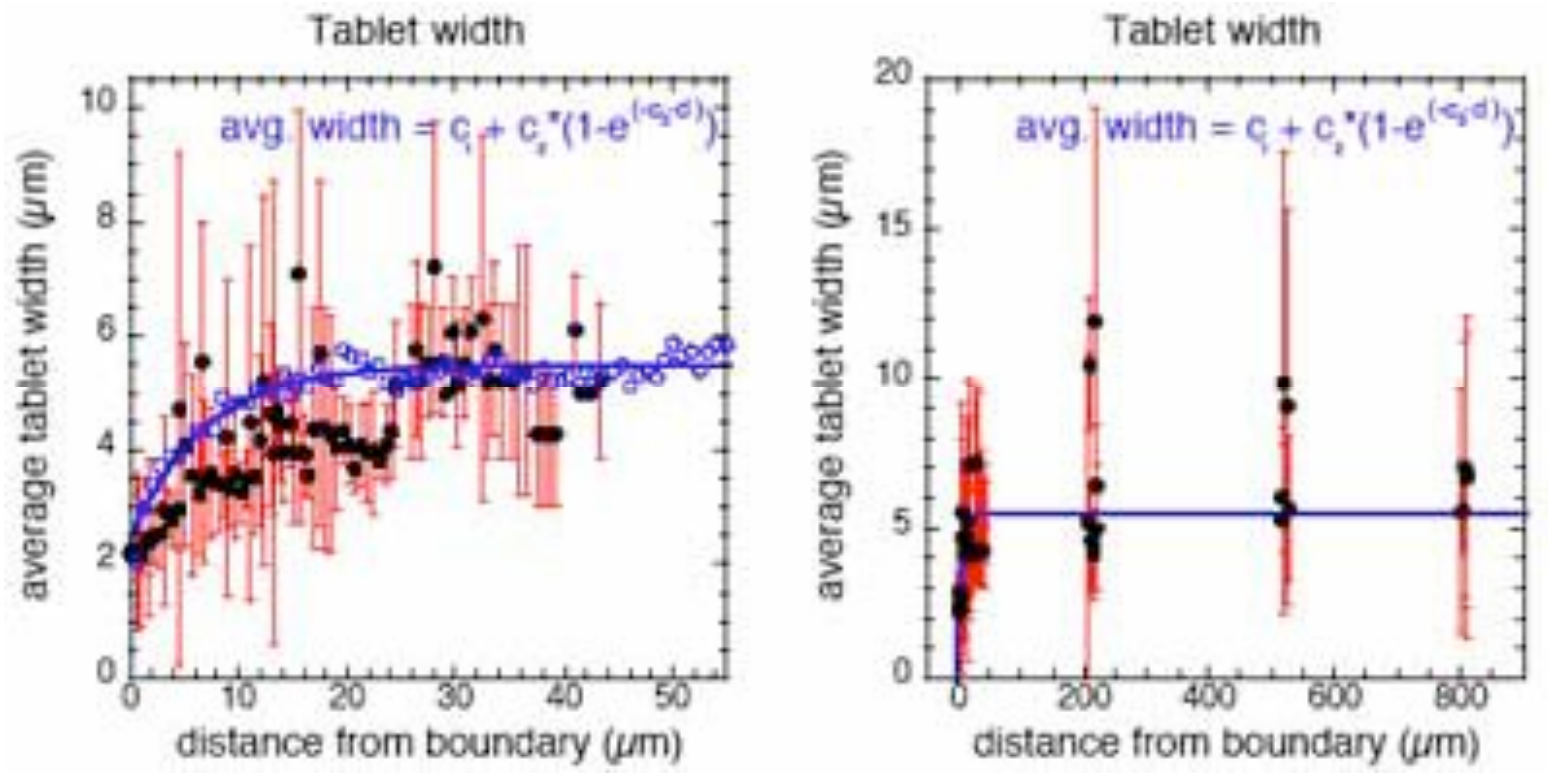

Supplementary Figure 1. Evolution of aragonite tablet widths with distance from the prismatic-nacre boundary. **(A)** The average tablet width versus distance from the nacre-prismatic boundary. The error bars represent the standard deviation. The widths of 1500 tablets from five different regions in three different shells were measured. The exponential fit shown (blue curve) is obtained by fitting the exponential form to the results of a numerical simulation of our model described in Supplementary Text Section 1b with parameter values ε=0.025, δ=0.3 (the resulting fit parameters $c_1$, $c_2$ and $c_3$ are 2.47 μm, 3.06 μm and 0.15 μm$^{-1}$; in the fit formula, d is the distance from the nacre-prismatic boundary). (**B**) The average tablet width in a single shell was measured up to a distance of 800 μm, and found to vary between 2-20 μm. Tablets reach their final average width of 5.5 μm at distances greater than 50 μm from the boundary.

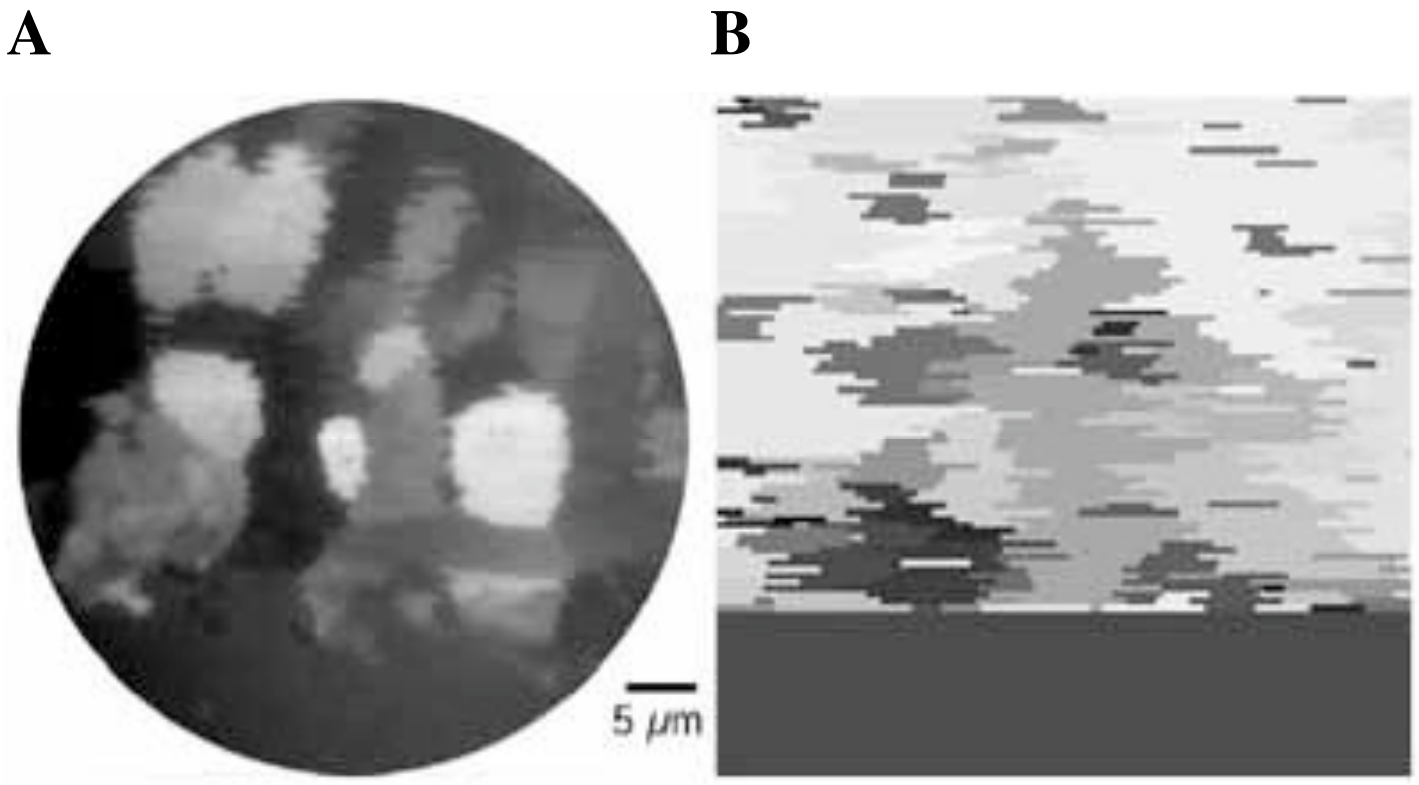

Supplementary Figure 2. Variability of tablet stacking morphology from experimental measurements on nacre and from the theoretical model. Experimental data and theoretical model of nacre growth near the prismatic-nacre boundary. In both images from bottom to top we display the prismatic layer, the boundary, and nacre. (**A**) Among the dozens of regions at the boundary we analyzed in red abalone nacre, this is the only one exhibiting larger tablet width near the boundary. (**B**) Simulation results for a nacre model in which all input parameters were identical to those of Figure 3B (ε=0.028, δ=0.2) but the tablets appear considerably wider. The similarity of tablet width in these two unusual cases shows that statistical variability is sufficient to justify the anomalous tablet width seen in this nacre region.

**Supplementary Text Section 1: Properties of the model incorporating differential growth rates and co-oriented nucleation.**

In our model of nacre growth, the tablets in each layer are nucleated at random locations, and it is assumed that growth in a given layer is completed before tablets in the succeeding layer are nucleated. The growth rates in the first layer are chosen uniformly at random in the interval [1-δ/2, 1+δ/2], and the tablets in each layer grow to confluence. With probability 1-ε a tablet has the same growth rate as the tablet below its nucleation site, while with probability ε the tablet is assigned a growth rate chosen uniformly at random from the range [1-δ/2, 1+δ/2]. Figure 2 and Supplementary Figure 1 show that the experimentally measured dependence of tablet widths and gray levels on position match closely those of the model for the parameter values ε=0.025, δ=0.3, while the pattern of nacre tablet orientations on a single sample shown in Figure 3 matches closely the results of a numerical simulation of this model using parameter values ε=0.028, δ=0.2. Here, we illustrate the dependence of the behavior on the parameters ε and δ. Section b) presents the details of the method used to compare the model to the experiment, and Section c) presents a simple analytic probabilistic analytic model that enables one to understand qualitatively some features of the behavior. The analytic model is closely related to models studied in population biology *(S1-S5)*.

**a) Numerical results.**
Simulation method: Each layer in the simulation was a square of dimension 400×400 (the length unit is arbitrary, and is adjusted when the results are compared with experimental data) with 600 nucleation sites randomly chosen from a probability distribution with uniform spatial density. The growth rates in the initial layer are chosen uniformly at random in the interval [1-δ/2, 1+δ/2]. In subsequent layers, 600 nucleation points are randomly chosen with uniform spatial density in the 400×400 square. For each nucleation site, with probability 1-ε the tablet's growth rate is the same as for the tablet directly below the nucleation site, and with probability ε it is chosen randomly and uniformly in the range [1-δ/2, 1+δ/2]. The outer 10% of the transverse dimensions of the simulational domain are discarded in all plots to avoid possible boundary effects.

The main text shows the results of a numerical simulation with parameter values δ=0.2 and ε=0.028. Here, runs with different values of δ and ε are shown to illustrate how the behavior depends on the choice of model parameters. Supplementary Figure 3 shows runs with ε=0 and two values of δ, 0.20 and 0.30.

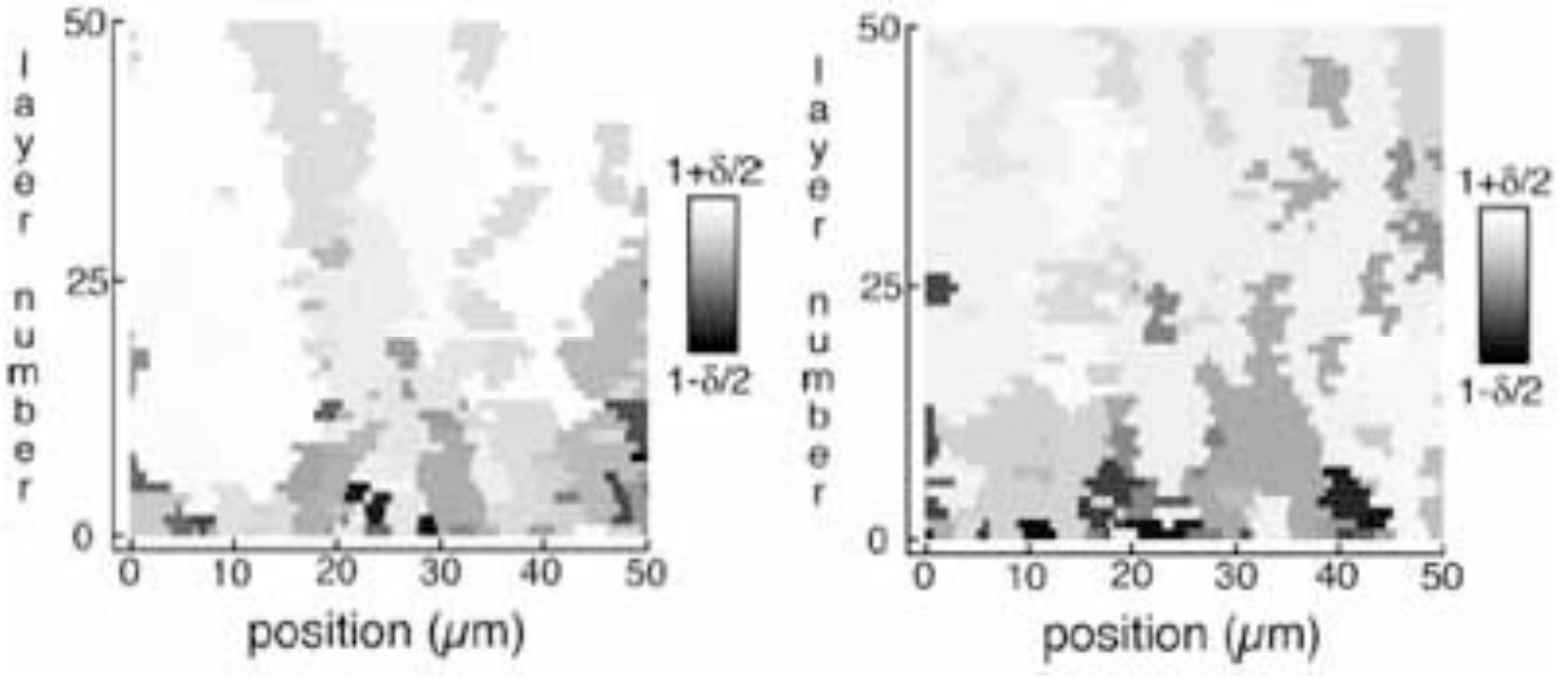

Supplementary Figure 3: Results of model simulations with different values of the parameter δ governing the growth rate variability. Configurations of the model with ε=0 (all tablets have the orientation of tablet below their nucleation sites). The graphs are two-dimensional vertical cuts through a three-dimensional sample. Left panel: δ=0.2, and right panel: δ=0.3. The gray scales in the figures show the growth rate range [1-δ/2,1+δ/2], with lighter colors denoting larger values. Changing δ changes only the rate at which ordering evolves.

Supplementary Figure 4, in which we present configurations of the model with δ=0.2 and two values of ε, 0.05 and 0.1, shows that increasing ε decreases the tablet stack heights and also cuts off the development of the ordering.

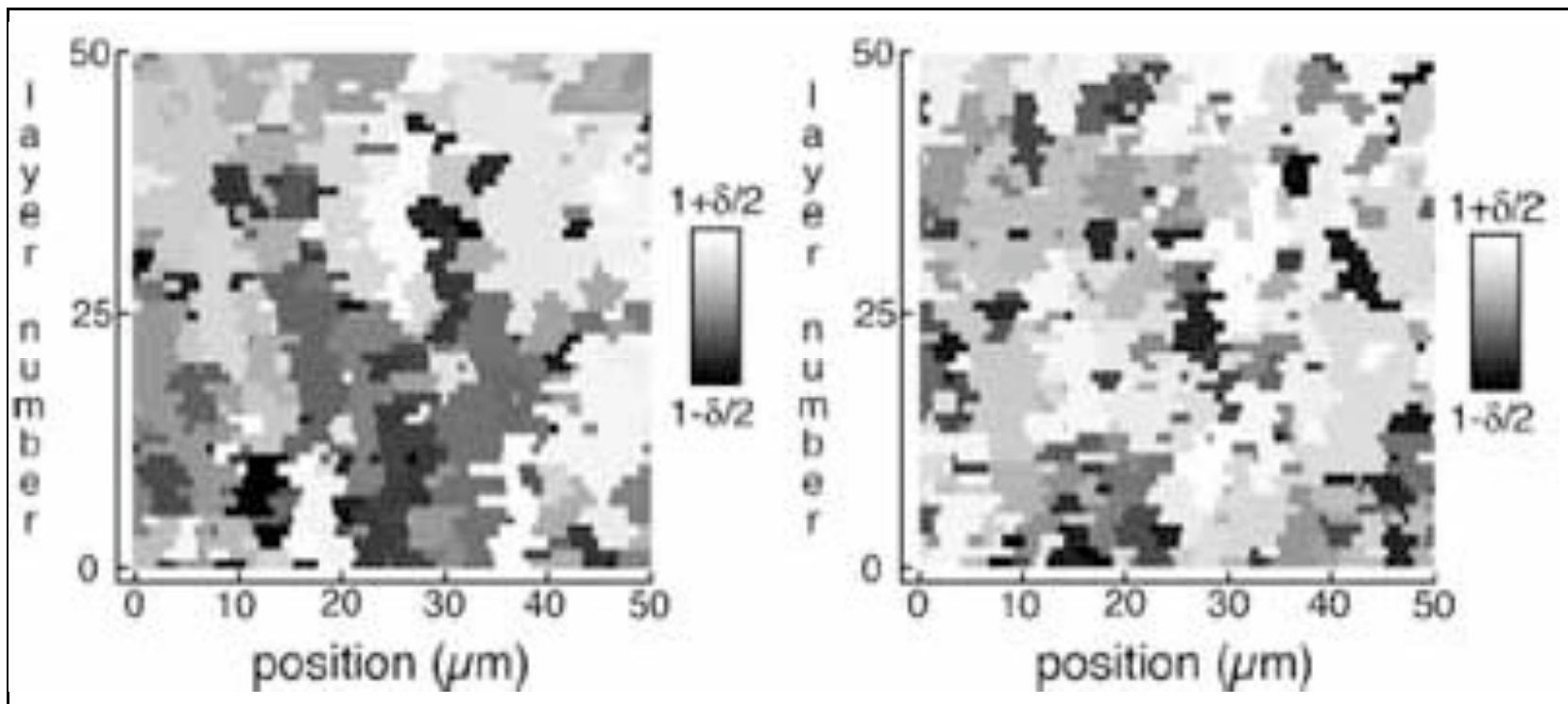

Supplementary Figure 4: Results of model simulations with different values of the parameter ε governing the fraction of tablets, which fail to co-orient with the tablet below their nucleation site. Two-dimensional slices of three-dimensional configurations of the model with δ=0.2. Left panel: ε=0.05, and right panel: ε=0.1. The gray scales in the figures show the range [1-δ/2,1+δ/2], with lighter colors denoting larger values. Increasing ε decreases the size of the tablet stacks and cuts off the development of the ordering, so that the degree of order saturates instead of growing indefinitely as when ε=0 (Supplementary Figure 3).

**b) Procedure for comparing model to experimental data.**
The initial conditions for the model were chosen to correspond to a uniform distribution of initial tablet orientations (the solid angle corresponding to a range of values between γ and γ+dγ between the tablet c-axis and the normal to layers is proportional to sin γ dγ = d(cos γ), while the tablet growth rate as a function of misorientation γ varies quadratically near its maximum, and so can be approximated as v cos γ). The parameter δ that characterizes the range of growth rates depends both on the range of initial tablet misorientations and on the magnitude of the variation of the growth velocity with misorientation angle.

The values for tablet widths were taken to be $\frac{1}{\sqrt{N_t(\ell)}}$, where $N_t(\ell)$ is the number of tablets with distinct growth rates in layer $\ell$. The conversion between number of layers and distance is straightforward, using the layer thickness that we measured to be 0.4 μm. The one unknown overall scale parameter of the tablet widths is adjusted by matching the model results to experimental data. The $\ell$-dependence of $N_t(\ell)$ is insensitive to the value of δ unless it is quite large; the size of the overall change in the mean tablet width increases as ε is decreased, while the initial rise in the mean tablet width is determined by the effective random walk of tablet boundaries and is not sensitive to the values of either of the parameters in the model.

The experimental data for the difference between the minimum and maximum gray scale values for tablets in a given layer (mean tablet widths ~6 μm) in a field of view of ~40 μm were compared to the values of the 15$^{th}$ percentile of the gray scale values in the model at the nucleation sites in each layer. Once again, the conversion from number of layers to distance is known, and an overall scale factor for the gray scale is determined by comparing to the experimental data.

**c) Analytic theory.**

The theory presented here is similar to Ginzburg-Landau theories (*S6*) commonly used in statistical physics (*S7*) in that the functional forms follow from symmetry considerations and therefore they involve some unknown coefficients. Even though the numerical values of some coefficients are not known, the analytic theory is very useful for obtaining insight into the interplay between the various parameters in the problem. The theory is closely related to other theories that have been studied in the context of population biology (*S1-S5,S8*).

It is useful to think explicitly about the analogy between the growth model for nacre and the mutation-selection models used in population biology. Our model posits that variability in the orientations of the c-axes of the tablets gives rise to variability in the tablet growth rates, and moreover that the tablets with c-axes perpendicular to the layer grow the fastest. The faster growth of tablets of a given orientation is analogous, in population biology, to species with higher fitness that reproduce faster than species with lower fitness. When a tablet in the next layer nucleates, one of two things happen. The first possibility is for the tablet to have the same orientation as the tablet directly below its nucleation site (analogous to inheritance), and the second is for the tablet to have a randomly chosen orientation (analogous to mutation). Selection is the tendency to get a larger fraction of the area filled with tablets with higher growth rates, which occurs because tablets that grow quickly will take up more of the area than those that grow slowly, increasing the probability that a given randomly located nucleation point is over a tablet with a higher growth rate.

We define $\phi_\ell(\gamma)$ to be the fraction of the area in layer $\ell$ misoriented from the normal by an angle $\gamma$. This function is a direct analog of a fitness function in population biology. Tablets with $\gamma = 0$ grow fastest and therefore their share of the area in layer $\ell+1$ will be greater than in layer $\ell$. Because the maximum is at $\gamma = 0$, one expects the dependence of $\phi_\ell(\gamma)$ on $\gamma$ to be quadratic. Therefore,

$$\frac{\phi_{\ell+1}(\gamma)}{\phi_\ell(\gamma)} = \frac{1}{N_{\ell+1}}\left(1 - \frac{\alpha\gamma^2}{2}\right) \approx \frac{1}{N_{\ell+1}} e^{-\alpha\gamma^2/2},$$

where $\alpha$ is a numerical coefficient and $N_\ell$ is a normalization factor that is fixed by the normalization condition

$$1 = \int_{-\frac{\pi}{2}}^{\frac{\pi}{2}} \phi_{\ell+1}(\gamma)d\gamma = \frac{1}{N_\ell}\int_{-\frac{\pi}{2}}^{\frac{\pi}{2}} e^{-\alpha\gamma^2/2}\phi_\ell(\gamma)d\gamma.$$

We assume that the distribution of angles $\gamma$ is not too broad (an assumption that is well-satisfied in nacre), so that replacing the quadratic with a Gaussian is a good approximation, and also that extending the limits of integration from $-\infty$ to $\infty$ introduces negligible error, and find

$$N_{\ell+1} = \int_{-\infty}^{\infty} e^{-\alpha\gamma^2/2}\phi_\ell(\gamma)d\gamma.$$

We also assume that there is a small probability $\varepsilon$ that, when a tablet nucleates, instead of adopting the same $\gamma$ as the tablet below its nucleation site, it takes on a value chosen from a probability distribution $w(\gamma)$, that we will choose that to also be Gaussian, $w(\gamma)=(\beta/2\pi)^{1/2}\exp(-\beta\gamma^2/2)$. (This process is analogous to a mutation term in a population genetics model (*S3*).) Therefore, the overall evolution is governed by the dynamical equation:

$$\phi_{\ell+1}(\gamma) = \varepsilon\left(\frac{\beta}{2\pi}\right)^{1/2} e^{-\beta\gamma^2/2} + (1-\varepsilon)\frac{1}{N_{\ell+1}}\phi_\ell(\gamma)e^{-\alpha\gamma^2/2}.$$

First note that when the co-orientation of tablets in successive layers is perfect, so that $\varepsilon=0$, this model is easily solved for any initial distribution, $\phi_1(\gamma)$, by noting that the normalization at each step need not be calculated explicitly, so that

$$\phi_{\ell+1}(\gamma) \propto \phi_1(\gamma)e^{-\ell\alpha\gamma^2/2}.$$

It follows immediately that when the co-orientation of successive layers is perfect, then for any $\alpha > 0$ the width of the distribution decreases as the square root of the number of layers and becomes arbitrarily narrow as the number of layers tends to infinity.

Now we consider the effects of nonzero but small $\varepsilon$, so that the co-orientation between successive layers is not quite perfect. The intuitive picture of the process is that the distribution gets narrower unless a misorientation, or “mutation,” occurs. The “mutations” prevent the peak from narrowing indefinitely, so after many layers $\phi_\ell(\gamma)$ approaches a stationary distribution that does not change as $\ell$ increases further. Since a fraction $\varepsilon$ mutates at each layer, the peak in the distribution gets narrower for $\sim(1/\varepsilon)$ steps. The width of the peak in $\phi(\gamma)$ after a peak that starts with width $(\beta)^{-1/2}$ has been narrowed for $1/\varepsilon$ steps is of order $(\beta+\alpha/\varepsilon)^{-1/2}$ When $\alpha/\varepsilon \gg \beta$, this width is of order $(\varepsilon/\alpha)^{1/2}$. A full analysis using the techniques in (*S5*) confirms this simple picture. The result implies that the degree of ordering is limited by the reliability of the co-orientation of crystals in successive layers.

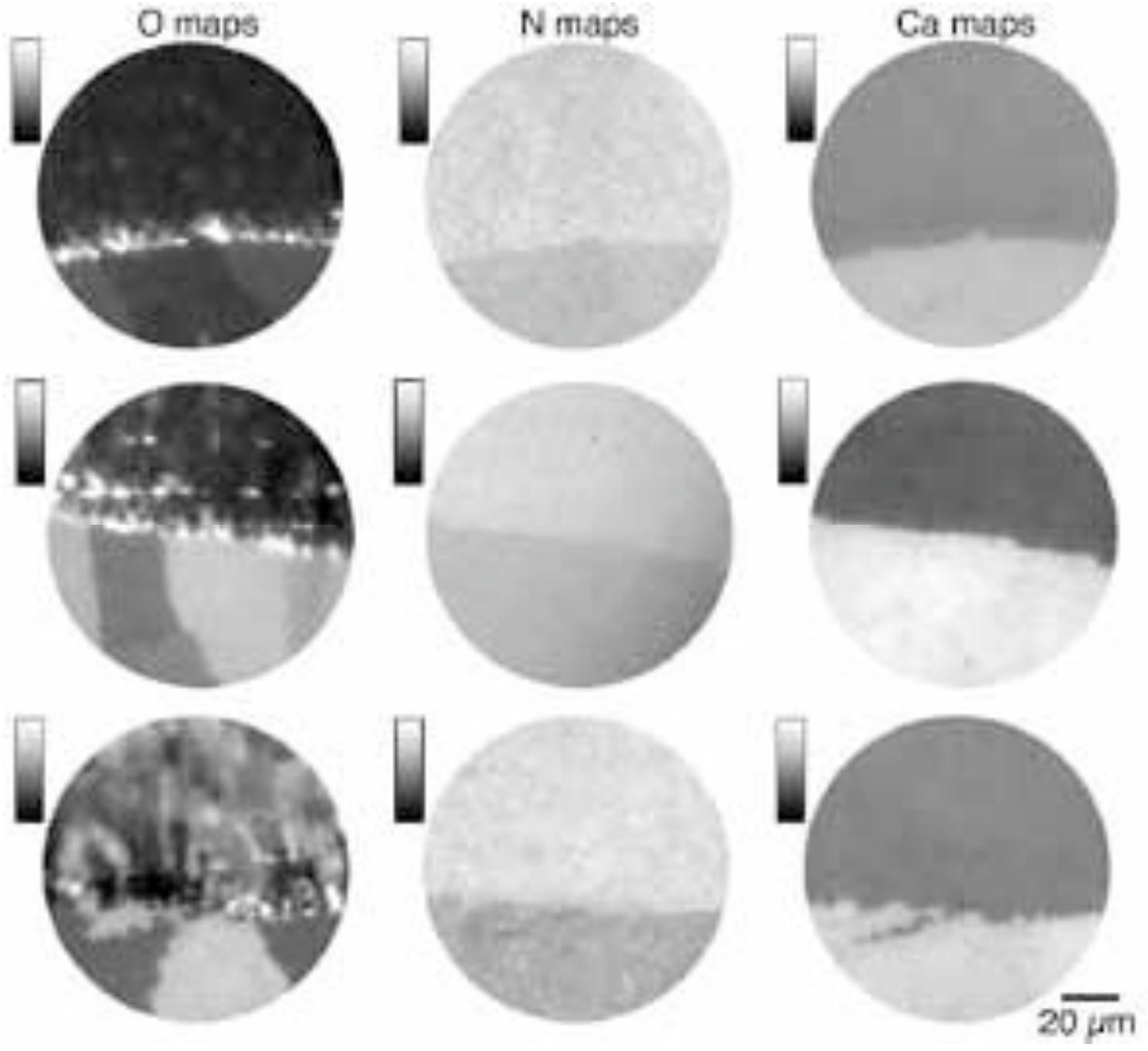

Supplementary Figure 5. Oxygen, nitrogen, and calcium maps. Oxygen, Nitrogen and Ca distribution maps in three different regions at the prismatic-nacre boundary in red abalone nacre. The O maps were obtained by digital ratio of images at 534 and 530 eV, the N maps by ratio of 402 and 394 eV images (N1s→π* peak and pre-edge, respectively), and the Ca maps by ratio of 351 and 344 eV images. The O maps show crystal orientation-dependent PIC both in nacre (top of each image) and prismatic layer (bottom). The N maps show higher N concentration (brighter pixels) in nacre compared to prismatic layer. The Ca maps were not acquired on one of the main Ca peaks, but on a crystal field peak at 351 eV, which is very prominent in calcite and not in aragonite. While the overall concentration of Ca is similar in nacre and prismatic layers, the calcite peak is most intense in the prismatic layer, thus these maps appear brighter in the prismatic than in the nacreous layer. Notice how the N concentration is homogenous in the nacre layer, and does not decay with distance from the prismatic-nacre boundary. The N concentration captures the overall protein concentration, and does not distinguish that of the ordering protein controlling aragonite (001) growth rate. We found no evidence for more concentrated proteins at the boundary. Based on negative evidence, therefore, it is plausible that the ordering protein is also homogeneously distributed in nacre at and away from the boundary.

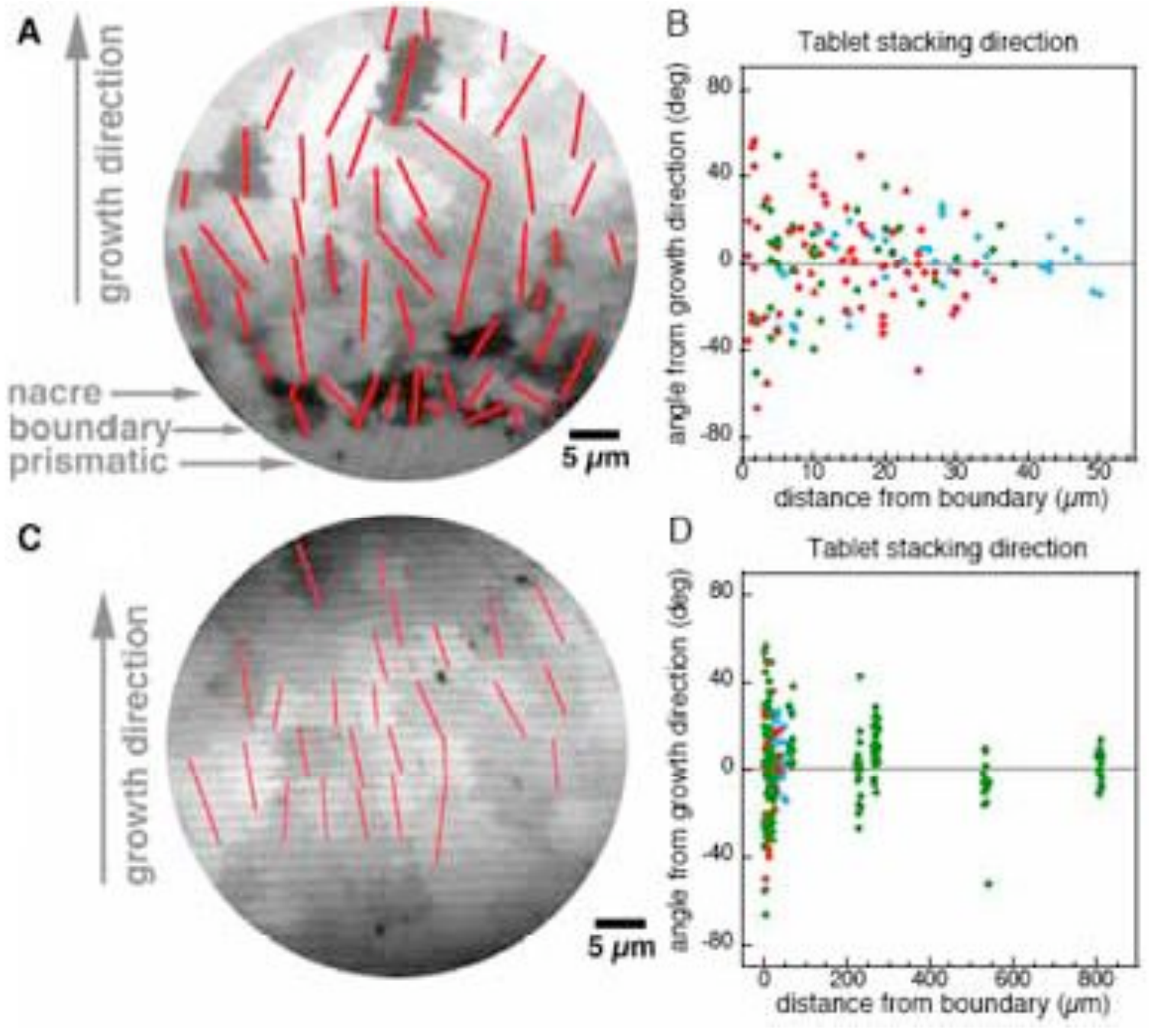

Supplementary Figure 6. Evolution of degree of order in tablet stacking direction with distance from the prismatic-nacre boundary. Tablet stacking direction as a function of distance from the boundary. (**A**) Oxygen π*/pre-edge map of 60° section of red abalone shell at the boundary. The axis of each tablet stack was drawn in red over the image, to indicate the direction of stacking with respect to the nacre growth direction, which in this image is vertical and upward. (**B**) All measured angles between the red axes and the growth direction are plotted as a function of distance from the boundary, corrected for polishing angles. The distance was measured at the center of each axis. Stack directions from three different shells were measured, and are reported in different colored dots. Notice how the angles are comprised within ±60° at the boundary and only within ±20° after 30 µm from the boundary. (**C**) Oxygen π*/pre-edge map of the same red abalone sample as in (**A**), analyzed approximately 300 µm from the boundary. (**D**) The angles measured from the map in (**C**) and others, show a variation within ±20° far from the boundary. These angles are underestimated for two distinct reasons. The sample surface shows a 60° section rather than a 90° cross-section. The layers appear therefore thicker, and the axis angles are smaller. Furthermore, since we are looking at a 2D section and not a 3D representation of nacre, the components of all angles towards and away from the reader are not apparent. Both effects contribute to an underestimation of the stacking direction angles. It is possible, therefore, that the real angles at the boundary vary by more than ±60°, even possibly approaching the limit of randomly directed stacks, which vary between ±90°. It is remarkable how, from the complete disorder of random or near-random directions, stacking of co-oriented tablets is harnessed within a mere 50 µm. The model also yields stacking angles that decrease as one moves away from the boundary even for tablet growth that is fully layer-by-layer and from random nucleation sites in each layer, because the mean tablet width is larger farther from the prismatic boundary. However, the degree of orientational ordering depends strongly on whether or not the growth is completely layer-by-layer, and detailed comparison of tablet orientations with models with layer growth that is not fully sequential is an interesting open problem.

## Supplementary References